\documentclass[twocolumn,showpacs,prb]{revtex4}
\usepackage{graphicx}
\usepackage{bm}

\begin{document}

\title{Orbital order in the low-dimensional quantum spin system TiOCl probed by ESR}
\author{V.~Kataev$^{1,2}$}\email{kataev@ph2.uni-koeln.de}
\author{J.~Baier$^{1}$}
\author{A. M\"{o}ller$^{3}$}
\author{L.~Jongen$^{3}$}
\author{G.~Meyer$^{3}$}
\author{A.~Freimuth$^{1}$}
\affiliation{$^{1}$ II. Physikalisches Institut, Universit{\"{a}}t zu K{\"{o}}ln,
                50937 K{\"{o}}ln, Germany}
\affiliation{${^2}$ Kazan Physical Technical Institute, Russian Academy of
Sciences, 420111 Kazan, Russia}
\affiliation{$^{3}$ Institut f\"{u}r Anorganische Chemie, Universit\"{a}t zu K\"{o}ln, 50939
K\"{o}ln, Germany}

\date{\today}

\begin{abstract}
We present electron spin resonance data of Ti$^{3+}$ (3$d^1$) ions in single crystals of the novel
layered quantum spin magnet TiOCl. The analysis of the $g$ tensor yields direct evidence that the
$d_{xy}$ orbital from the $t_{2g}$ set is predominantly occupied and owing to the occurrence of
orbital order a linear spin chain forms along the crystallographic $b$ axis. This result
corroborates recent theoretical LDA+U calculations of the band structure. The temperature
dependence of the parameters of the resonance signal suggests a strong coupling between spin and
lattice degrees of freedom and gives evidence for a transition to a nonmagnetic ground state at
67~K.
\end{abstract}


\maketitle

\begin{figure}
\includegraphics[angle=0,width=0.8\columnwidth]{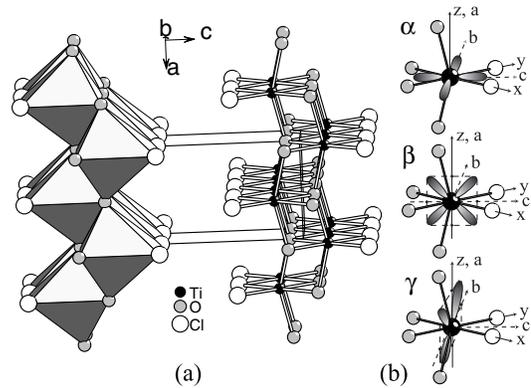}
\caption{(a) Crystal structure of TiOCl; (b) sketch of the $t_{2g}$ orbitals
($\alpha$, $\beta$ and $\gamma$) in the [TiO$_4$Cl$_2$] octahedron. The local
coordinate frame \{xyz\} is chosen so that z$\parallel\!a$ axis, and the $x$ and $y$
axes are rotated by  45$^\circ$ with respect to the $c$ and $b$ axes, respectively.
Note that in the LDA+U calculation of Seidel {\it et al.}~\protect\cite{Seidel}
one-dimensional bands are formed from overlapping $xy$, $yz$, and $xz$ orbitals,
respectively, whereas on a single cluster level the $\alpha\!=\!xy$,
$\beta\!=\!(xz\!+\!yz)/\sqrt{2}$ and $\gamma\!=\! (xz\!-\!yz)/\sqrt{2}$ states are the
eigenstate orbitals, respectively.}
\label{structure}
\end{figure}

Transition metal (TM) oxides with low-dimensional structural
elements provide a fascinating 'playground' to study novel
phenomena  such as high-temperature superconductivity,
spin-charge separation, spin-gap states and quantum disorder
\cite{Orenstein,Dagotto,Sachdev}. Until recently, the emphasis
has been put on Cu-based oxides, where a Cu$^{2+}$ (3$d^9$) ion
has a single hole in the $e_g$ orbitals with spin $S\!=\!1/2$,
and its orbital momentum is almost completely quenched by the
crystal field. The ions at the beginning of the TM elements
row, like Ti$^{3+}$ and V$^{4+}$, have, in contrast, a single
$d$-electron which occupies one of the $t_{2g}$ orbitals.
Because these orbitals are much less Jahn-Teller active, their
near degeneracy may yield more complicated physics, involving
not only the spin and charge, but also the orbital sector
\cite{Tokura}. As an example, the three-dimensional cubic
perovskite  LaTiO$_3$ has been proposed to realize a quantum
orbital liquid \cite{Keimer,Khaliullin}. However recent x-ray
and neutron structural data suggest the ordering of the
orbitals \cite{Cwik}. The structural dimensionality is reduced
in TiOCl, where [TiO$_4$Cl$_2$] octahedra are arranged in
bilayers  separated from each other along the $c$ axis
(Fig.~\ref{structure}a). In fact, for quite a while this
compound has been considered as a 2D antiferromagnet, an
electron analog to the high-$T_c$ cuprates \cite{Beynon}, owing
to an almost $T$-independent magnetic susceptibility reported
in Ref.~\onlinecite{Maule}. However, very recently TiOCl has
emerged in an entirely new light as a 1D antiferromagnet
\cite{Seidel} and is proposed as the second example of an
inorganic spin-Peierls compound after CuGeO$_3$ \cite{Hase}.
LDA+U band-structure calculations \cite{Seidel} suggest
ordering of the $t_{2g}$ orbitals in TiOCl which produces
quasi-1D antiferromagnetic (AF)  $S\!=\!1/2$ chains. This
calculation favors the occupancy of the $d_{xy}$ orbitals
(Fig.~\ref{structure}b) which form a uniform chain along the
$b$ axis. A transition to a non-magnetic state at
$T_c\!=\!67$~K has been observed in the static magnetization
\cite{Seidel}. Remarkably, NMR data reveal the pre-existing
pseudo spin-gap already above $T_c$ which is ascribed to strong
orbital fluctuations \cite{Imai}.

In this paper we present electron spin resonance (ESR) data of Ti$^{3+}$ (3$d^1$) in
single crystals of TiOCl. By measuring the anisotropy of the $g$ factor and comparing
it with our theoretical estimates in the framework of the angular overlap model we
conclude that the  single $d$ electron occupies the $d_{xy}$ orbital which lies in the
$bc$ plane. This result suggests a formation of a spin-1/2 chain along the $b$
direction, owing to the overlap of the orbital states, and supports recent LDA+U
calculations \cite{Seidel}. The ESR signal vanishes at $T_c\!=\!67$~K signalling the
transition to a non-magnetic ground state. A pronounced dependence of the linewidth
and the $g$ factor on temperature suggests a strong coupling of spins to the lattice
which may play an important role for the opening of the spin gap.

\begin{figure}
\includegraphics[angle=0,width=0.9\columnwidth]{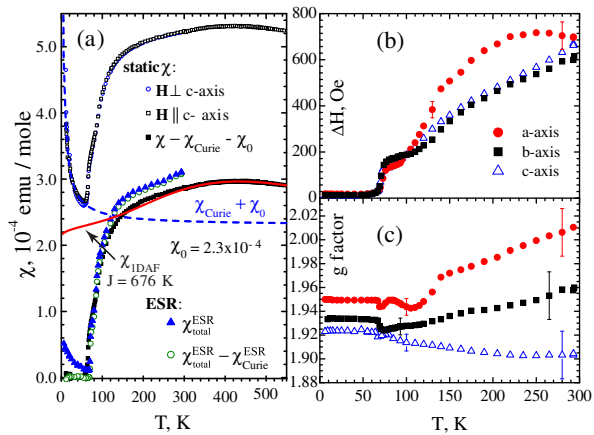}
\caption{(a) Static susceptibility $\chi(T)$ of TiOCl together with the ESR spin susceptibility and
corresponding fits (see text); (b) and (c) ESR linewidth $\Delta H$ and $g$ factor, respectively.}
\label{chiESRvsT}
\end{figure}

Single crystals of TiOCl have been prepared from TiCl$_3$ (Aldrich) and TiO$_2$
(Kronos Titan) according to experimental details given in Ref. \onlinecite{Schaefer}.
The purity of the product was checked by x-ray powder diffraction at 293~K and 10~K.
The latter measurement has been carried out in order to check for a possible
structural phase transition, which we do not find \cite{commentdimer}. Both
diffractograms could be indexed in the orthorhombic space group {\it Pmmn} with
$a\!=\!3.789(1)$ \AA, $b\!=\!3.365(1)$ \AA, $c\!=\!8.060(3)$\AA\ at 293~K and
$a\!=\!3.7946(3)$ \AA, $b\!=\!3.3584(2)$ \AA, $c\!=\!8.057(1)$~\AA\ at 10~K. It is
interesting to note that the $a$ axis is longer and the $b$ and $c$ axes are shorter
at lower temperatures, which means that the already distorted [TiO$_4$Cl$_2$]
octahedra are slightly less compressed at 10~K. The static magnetic susceptibility
$\chi(T)$ has been measured in a Faraday balance in a field of 1~Tesla in the
temperature range $10\!-\!550$~K (Fig.~\ref{chiESRvsT}a). Within the experimental
accuracy $\chi(T)$ is isotropic and shows a behavior similar to that reported in
Ref.~\onlinecite{Seidel}. The fit of the data with the formula
$\chi(T)\!=\!\chi(T)_{\rm 1D AF}\!+\!\chi(T)_{\rm Curie}\!+\!\chi_0$, where
$\chi(T)_{\rm 1D AF}$ is the susceptibility of a uniform 1D Heisenberg antiferromagnet
\cite{Klumper}, $\chi(T)_{\rm Curie}$ is the Curie term, and $\chi_0$ is the sum of
the diamagnetic susceptibility of TiOCl and the Van-Vleck susceptibility of Ti$^{3+}$,
yields the number of free $S\!=\!1/2$ spins (paramagnetic defects), responsible for
the low-$T$ Curie upturn of $\chi(T)$ of order 0.6 \%, and
$\chi_0\!=\!2.3\!\times\!10^{-4}$ emu/mole \cite{notechi0}. After the subtraction of
$\chi(T)_{\rm Curie}$ and $\chi_0$ from the raw data one sees that the model of a 1D
antiferromagnet with the nearest neighbor exchange $J\!=\!676$~K and the $g$ factor of
1.91 (see below) describes the data in the high-$T$ regime quite well
(Fig.~\ref{chiESRvsT}a). Below 130~K the susceptibility rapidly decreases and
approaches zero at $\sim 60$~K \cite{notechigap,frustration,Dumoulin}.

ESR has been measured using a Bruker spectrometer at X-band frequency 9.48 GHz and at
temperatures between 2 and 300~K. A single resonance line of a Lorentzian shape has
been observed. The intensity of the ESR signal $I$ is proportional to the
susceptibility of the resonating spins \cite{Abragam}. A comparison of $I$ of TiOCl
with that of a reference sample \cite{ruby} gives evidence that within the
experimental uncertainty practically all spins contributing to the static
susceptibility participate in ESR. The $T$-dependence of the spin susceptibility
$\chi^{\rm ESR}$ obtained from the integrated intensity $I(T)$ is similar to that of
the static $\chi$ (Fig.~\ref{chiESRvsT}a) \cite{noteintensity}. In particular,
$\chi^{\rm ESR}$ drops sharply almost to zero at $T_c\!=\! 67$~K signalling the
transition to a non-magnetic state. In addition a small kink can be seen at
$90\!-\!95$ K, which is ascribed in Ref.~\onlinecite{Imai} to the onset of the
structural instability. A small signal arising at $T\!<\!T_c$ is obviously due to a
small amount of paramagnetic impurities in the samples (e.g. Ti$^{3+}$ in structural
defects). The bulk signal observed above $T_c$ is anisotropic and exhibits a strong
temperature dependence of the $g$ factor and the linewidth $\Delta H$
(Fig.~\ref{chiESRvsT}b and c).

\begin{figure}
\includegraphics[angle=-90,width=0.9\columnwidth]{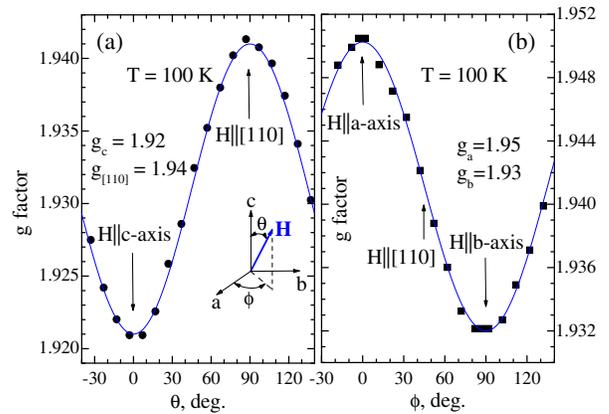}
\caption{Angular dependence of the $g$ factor at $T\!=\!100$ K:
(a) (1-10) plane; (b) (001) plane.} \label{ganisotropy}
\end{figure}

A representative example of the dependence of $g$ on the direction of the external
magnetic field $H$ is shown in Fig.~\ref{ganisotropy}. One notices that the $c$ axis
is a symmetry axis of the $g$ tensor, where $g$ reaches its minimum value
(Fig.~\ref{ganisotropy}a). The $g$ factor in the $ab$ plane is larger and also
anisotropic, as expected in the orthorhombic symmetry (Fig.~\ref{ganisotropy}b): The
situation is typical for a Ti octahedron strongly compressed along the $z$ axis with
smaller distortions in the equatorial plane, where the $xy$ orbital ground state is
realized \cite{Pilbrow}. On the quantitative level we compare the experimental data
with the results of the angular overlap model (AOM). The AOM has been a tool to
rationalize spectroscopic and magnetic properties of TM and rare-earth metal ions in
various ligand fields and provides a mathematical approach to calculating relative
energies of molecular orbitals of a TM complex from the overlap of the central-atom
orbitals with the ligand orbitals \cite{Jorgensen}. We calculated the $g$ values for
TiOCl using the program package CAMMAG \cite{Cruse,AOMdetails}. The best fit of the
$g$ tensor with experimental data (Table~\ref{gAOM}) clearly identifies the $d_{xy}$
orbital as the ground state orbital. The relative energies of the orbital states and
the bonding parameters used in the calculation are listed in Table~\ref{AOMparam}.
Even if one neglects a plausible anisotropy of the $\pi$ interactions (Model A) the
AOM predicts correctly a minimum value of the $g$ factor along the $c$ axis and larger
values in the $ab$ plane ($g_c\!<\!g_b<\!g_a$). The agreement becomes even
quantitative if one accounts for the anisotropy of the $\pi$ bonding, which is quite
reasonable owing to the fact the ligands are all bridging, e.g.
[TiO$_{4/4}$Cl$_{2/2}$] (Model B) \cite{comment_reduction}. In the LDA+U calculation
of Seidel {\it et al.} \cite{Seidel} the two higher-energy 1D bands are derived from
unmixed $d_{yz}$ and $d_{xz}$ orbitals, which both make an angle of $\sim 45^\circ$
with the $c$ axis and the $ab$ plane. Their occupation by the $d$ electron can be
excluded, as the symmetry of these states is different from the symmetry of the $g$
tensor.

\begin{table}
\caption{Observed and calculated $g$ tensor at $T\!=\!293$ K.
(A - isotropic, B - anisotropic $\pi$ interactions).}
\begin{center}
\begin{ruledtabular}
\begin{tabular}{cccc}
 & experiment & Model A & Model B \\
 \hline
 $g_a$ & 2.01 & 1.946 & 1.976 \\
 $g_b$ & 1.96 & 1.935 & 1.959 \\
 $g_c$ & 1.91 & 1.926 & 1.911 \\
\end{tabular}
\end{ruledtabular}
\end{center}
\label{gAOM}
\end{table}

\begin{table}
\caption{AOM parameters for [TiO$_4$Cl$_2$] complex and relative energies of the $d$ orbital
states.}
\begin{center}
\begin{ruledtabular}
\begin{tabular}{lrl|lr}
$\zeta$&$\rm 125\, cm^{-1}$& & & \\ B&$\rm 700\, cm^{-1}$& & & \\ C&$\rm 2800\,
cm^{-1}$& & & \\ $e_\sigma$&$\rm 9500\, cm^{-1}$,&$d$(Ti-O) $\!=\! 1.95$ \AA & &  \\
$e_{\pi}$&$\rm 2400\, cm^{-1}$& &$z^2$ &$\rm 20769\, cm^{-1}$ \\ $e_\sigma$&$4900\,
cm^{-1}$,&$d$(Ti-O) $\!=\! 2.25$ \AA &${x^2\!-\!y^2}$ &$\rm 14891\, cm^{-1}$ \\
$e_{\pi}$&$\rm 1200\, cm^{-1}$& &($xz$+$yz$)/$\sqrt{2}$ &$\rm 5900\,cm^{-1}$ \\
$e_\sigma$&$\rm 5700\, cm^{-1},$&$d$(Ti-Cl) $\!=\! 2.37$ \AA &($xz$-$yz$)/$\sqrt{2}$
&$\rm 2456\,cm^{-1}$  \\ $e_{\pi}$&$\rm 900\, cm^{-1}$& &$xy$& $\rm 0\,cm^{-1}$ \\
$k_{x}$& 0.6\mbox{\ \ \ \ \ \ }&$k_{y}$,  $k_{z}$ \ \ \ 0.9 & & \\
\end{tabular}
\end{ruledtabular}
\end{center}
\label{AOMparam}
\end{table}

The above discussion strongly supports the scenario proposed in Ref.~\onlinecite{Seidel} on the
basis of the band structure results. The occupation of the $d_{xy}$ states favors the occurrence of
AF spin chains along the $b$ axis owing e.g. to the direct overlap of the $d_{xy}$ orbitals
(Fig.~\ref{structure}). The neighboring chains in the [TiO$_4$Cl$_2$]$_\infty^2$ bilayer are
shifted in a staggered fashion by $a/2$ and $b/2$ along the $a$ and $b$ axes, respectively
(Fig.~\ref{structure}a), so that the inter-chain exchange interaction is expected to be weak and
frustrated. Thus almost perfectly isolated spin-$1/2$ chains with a relatively strong AF exchange
$J\!\sim\!700$~K, as deduced from the $\chi$ data, can be realized. The spin chains undergo the
transition to a spin-gap state at $T_c=67$ K. However, NMR data \cite{Imai} indicate that this is
not a conventional spin-Peierls transition as observed e.g. in CuGeO$_3$~\cite{Hase} because of a
much larger spin gap $\Delta_{gap}$ as compared with $T_c$\ ($2\Delta_{gap}/k_BT_c\!\simeq\!13$)
\cite{commentdimer}. Probably additional orbital degrees of freedom play an important role in
TiOCl. Indeed, the components of the $g$ tensor exhibit a strong $T$ dependence
(Fig.~\ref{chiESRvsT}c), suggesting that the energy of the orbital states alters appreciably with
$T$. This is a rather unusual feature; for example, in CuGeO$_3$ as well as in NaV$_2$O$_5$ which
is another well-known low-D spin-gap TM oxide, the $g$ values do not change with $T$
\cite{Yamada,Lohmann}.

A considerable decrease of the $g$ anisotropy at low $T$ signals a reduction of the
distortion of the Ti complex, in agreement with our x-ray data. Such an appreciable
coupling of the spin to the lattice should affect the $T$ dependence of the ESR
linewidth as well. In a concentrated paramagnet spin-spin interactions produce a
finite ESR linewidth owing to the anisotropic part of the superexchange \cite{Abragam}
${\cal H}^\prime\! =\! \sum S_i A S_j$. This interaction yields a second moment of the
line $M_2\!\sim\!A\!\sim\!(\Delta g/g)^2J\;$. \cite{Moriya} Here $\Delta g$ is the
deviation of $g$ from the spin-only value of 2, and $J$ is the strength of the
isotropic Heisenberg exchange ${\cal H}\! =\! J \sum S_i S_j$ which in the 3D case
narrows the signal so that its width reduces to $\Delta H\!\sim\!M_2^2/J\;$.
\cite{Kubo,comment3DESR} If the susceptibility $\chi(T)$ deviates from the Curie law,
$M_2$ may acquire a $T$ dependence proportional to $\chi_{\rm Curie}/\chi(T)\;$.
\cite{Huber} Taking $\Delta g/g\!\sim\!0.05$ and $J\!\simeq\!680$~K we obtain $\Delta
H$ of the order of 30 Oe. The much smaller value of this rough estimate as compared
with the experimental data for $T\!>\!T_c$ (Fig.~\ref{chiESRvsT}b) may be in part due
to the fact that it neglects the peculiarities of the bonding geometry in the 1D spin
chain which in certain cases may considerably boost $\Delta H$.
 \cite{Kataev} More serious is the discrepancy of the $T$ dependence of $\Delta H$. Because
$\chi_{\rm Curie}/\chi(T)$ and the average value of $\Delta g$ both increase with
lowering $T$, the width $\Delta H$ is expected to increase, too. However,
experimentally one finds the opposite behavior (Fig.~\ref{chiESRvsT}b), suggesting
that other mechanisms of spin relaxation e.g. via orbital and lattice degrees of
freedom have to be considered \cite{commentdH}. The interplay between orbital and spin
fluctuations in TiOCl has been proposed in the discussion of the NMR results
\cite{Imai}. It is argued that the opening of the pseudo-spin gap at $T^*\!\sim\!135$
K is related to the suppression of the spin-Peierls transition caused by fluctuations
of the orbital states. Remarkably, in the temperature interval $T_c\!<T<\!T^*$ the ESR
linewidth levels off at a minimum value of $\sim\!150\!-\!170$ Oe before dropping down
at $T_c$. This implies that the spin fluctuations in this temperature regime are
strongly suppressed as expected in the pseudo-gap regime. The spin dynamics recovers
above $T^*$ resulting in the increase of $\Delta H$. An additional $T$-dependent
contribution to $\Delta H$ could arise due to the spin-phonon coupling, which may be
significant as suggested by the strong sensitivity of the $g$ factor to the change of
the lattice parameters \cite{optics}.

In summary, we have studied electron spin resonance of Ti$^{3+}$ ions in single crystals of the
novel low-dimensional spin magnet TiOCl. The analysis of the $g$ tensor justifies the scenario that
in an apparently 2D structure uniform spin $S\!=\!1/2$ chains are formed along the $b$ direction.
The bulk ESR signal vanishes at $T_c\!=\!67$ K evidencing the transition to a nonmagnetic, possibly
a spin-Peierls state. The $T$ dependence of the $g$ values and the linewidth suggests that orbital
and lattice degrees of freedom may play a key role in the magnetic properties of TiOCl. In
particular, strong spin and probably also orbital fluctuation effects above $T_c$ may be
responsible for a peculiar temperature dependence of the ESR parameters.

We acknowledge useful discussions with M. Gr\"{u}ninger, T. Lorenz, D. Khomskii and S.
Streltzov, and thank P.~Lemmens for pointing our attention to this system. This work
was supported by the DFG through SFB 608. V.K. acknowledges support of the RAS through
project No. OFN03/032061/020703-996.

\end{document}